\RequirePackage{lineno}
\documentclass[prc,preprint,superscriptaddress,showpacs,amssymb,amsmath,amsfonts,aps,floatfix]{revtex4}
\usepackage{graphicx}
\usepackage{dcolumn}
\usepackage{bm}
\usepackage{epsfig}
  
\def\be{\begin{equation}}  
\def\ee{\end{equation}}  
\def\bea{\begin{eqnarray}}  
\def\eea{\end{eqnarray}}  
\def\tpI{$e p \rightarrow e \pi^+ n$ }

\def\tpV{$e p \rightarrow e \pi^+ (n)$ }

\def\tpIn{$e p \rightarrow e \pi^+ n$}

\def\tpVn{$e p \rightarrow e \pi^+ (n)$}

\def\mxepi{$M_x^{e\pi}$ }

\def\cthcm{$\cos(\theta^*)$}
\def\cthcmsp{$\cos(\theta^*)$ }
\def\cthcmn{\cos(\theta^*)}
\def\phicm{$\phi^*$}
\def\phicmsp{$\phi^*$ }

\def\>{\hskip .1in }

\begin{document}

\title{Target and Beam-Target Spin Asymmetries in Exclusive 
Pion Electroproduction for $Q^2>1$ GeV$^2$. I. $e p \rightarrow e \pi^+ n$} 

\newcommand*{\ANL}{Argonne National Laboratory, Argonne, Illinois 60439}
\newcommand*{\ANLindex}{1}
\affiliation{\ANL}
\newcommand*{\ASU}{Arizona State University, Tempe, Arizona 85287-1504}
\newcommand*{\ASUindex}{2}
\affiliation{\ASU}
\newcommand*{\CSUDH}{California State University, Dominguez Hills, Carson, CA 90747}
\newcommand*{\CSUDHindex}{3}
\affiliation{\CSUDH}
\newcommand*{\CMU}{Carnegie Mellon University, Pittsburgh, Pennsylvania 15213}
\newcommand*{\CMUindex}{4}
\affiliation{\CMU}
\newcommand*{\CUA}{Catholic University of America, Washington, D.C. 20064}
\newcommand*{\CUAindex}{5}
\affiliation{\CUA}
\newcommand*{\SACLAY}{CEA, Centre de Saclay, Irfu/Service de Physique Nucl\'eaire, 91191 Gif-sur-Yvette, France}
\newcommand*{\SACLAYindex}{6}
\affiliation{\SACLAY}
\newcommand*{\UCONN}{University of Connecticut, Storrs, Connecticut 06269}
\newcommand*{\UCONNindex}{7}
\affiliation{\UCONN}
\newcommand*{\FU}{Fairfield University, Fairfield CT 06824}
\newcommand*{\FUindex}{8}
\affiliation{\FU}
\newcommand*{\FIU}{Florida International University, Miami, Florida 33199}
\newcommand*{\FIUindex}{9}
\affiliation{\FIU}
\newcommand*{\FSU}{Florida State University, Tallahassee, Florida 32306}
\newcommand*{\FSUindex}{10}
\affiliation{\FSU}
\newcommand*{\GWUI}{The George Washington University, Washington, DC 20052}
\newcommand*{\GWUIindex}{11}
\affiliation{\GWUI}
\newcommand*{\ISU}{Idaho State University, Pocatello, Idaho 83209}
\newcommand*{\ISUindex}{12}
\affiliation{\ISU}
\newcommand*{\INFNFE}{INFN, Sezione di Ferrara, 44100 Ferrara, Italy}
\newcommand*{\INFNFEindex}{13}
\affiliation{\INFNFE}
\newcommand*{\INFNFR}{INFN, Laboratori Nazionali di Frascati, 00044 Frascati, Italy}
\newcommand*{\INFNFRindex}{14}
\affiliation{\INFNFR}
\newcommand*{\INFNGE}{INFN, Sezione di Genova, 16146 Genova, Italy}
\newcommand*{\INFNGEindex}{15}
\affiliation{\INFNGE}
\newcommand*{\INFNRO}{INFN, Sezione di Roma Tor Vergata, 00133 Rome, Italy}
\newcommand*{\INFNROindex}{16}
\affiliation{\INFNRO}
\newcommand*{\INFNTUR}{INFN, Sezione di Torino, 10125 Torino, Italy}
\newcommand*{\INFNTURindex}{17}
\affiliation{\INFNTUR}
\newcommand*{\ORSAY}{Institut de Physique Nucl\'eaire, CNRS/IN2P3 and Universit\'e Paris Sud, Orsay, France}
\newcommand*{\ORSAYindex}{18}
\affiliation{\ORSAY}
\newcommand*{\ITEP}{Institute of Theoretical and Experimental Physics, Moscow, 117259, Russia}
\newcommand*{\ITEPindex}{19}
\affiliation{\ITEP}
\newcommand*{\KNU}{Kyungpook National University, Daegu 41566, Republic of Korea}
\newcommand*{\KNUindex}{20}
\affiliation{\KNU}
\newcommand*{\LPSC}{LPSC, Universit\'e Grenoble-Alpes, CNRS/IN2P3, Grenoble, France}
\newcommand*{\LPSCindex}{21}
\affiliation{\LPSC}
\newcommand*{\MISS}{Mississippi State University, Mississippi State, MS 39762-5167}
\newcommand*{\MISSindex}{22}
\affiliation{\MISS}
\newcommand*{\UNH}{University of New Hampshire, Durham, New Hampshire 03824-3568}
\newcommand*{\UNHindex}{23}
\affiliation{\UNH}
\newcommand*{\NSU}{Norfolk State University, Norfolk, Virginia 23504}
\newcommand*{\NSUindex}{24}
\affiliation{\NSU}
\newcommand*{\OHIOU}{Ohio University, Athens, Ohio  45701}
\newcommand*{\OHIOUindex}{25}
\affiliation{\OHIOU}
\newcommand*{\ODU}{Old Dominion University, Norfolk, Virginia 23529}
\newcommand*{\ODUindex}{26}
\affiliation{\ODU}
\newcommand*{\URICH}{University of Richmond, Richmond, Virginia 23173}
\newcommand*{\URICHindex}{27}
\affiliation{\URICH}
\newcommand*{\ROMAII}{Universita' di Roma Tor Vergata, 00133 Rome Italy}
\newcommand*{\ROMAIIindex}{28}
\affiliation{\ROMAII}
\newcommand*{\MSU}{Skobeltsyn Institute of Nuclear Physics, Lomonosov Moscow State University, 119234 Moscow, Russia}
\newcommand*{\MSUindex}{29}
\affiliation{\MSU}
\newcommand*{\SCAROLINA}{University of South Carolina, Columbia, South Carolina 29208}
\newcommand*{\SCAROLINAindex}{30}
\affiliation{\SCAROLINA}
\newcommand*{\TEMPLE}{Temple University,  Philadelphia, PA 19122 }
\newcommand*{\TEMPLEindex}{31}
\affiliation{\TEMPLE}
\newcommand*{\JLAB}{Thomas Jefferson National Accelerator Facility, Newport News, Virginia 23606}
\newcommand*{\JLABindex}{32}
\affiliation{\JLAB}
\newcommand*{\UTFSM}{Universidad T\'{e}cnica Federico Santa Mar\'{i}a, Casilla 110-V Valpara\'{i}so, Chile}
\newcommand*{\UTFSMindex}{33}
\affiliation{\UTFSM}
\newcommand*{\EDINBURGH}{Edinburgh University, Edinburgh EH9 3JZ, United Kingdom}
\newcommand*{\EDINBURGHindex}{34}
\affiliation{\EDINBURGH}
\newcommand*{\GLASGOW}{University of Glasgow, Glasgow G12 8QQ, United Kingdom}
\newcommand*{\GLASGOWindex}{35}
\affiliation{\GLASGOW}
\newcommand*{\VIRGINIA}{University of Virginia, Charlottesville, Virginia 22901}
\newcommand*{\VIRGINIAindex}{36}
\affiliation{\VIRGINIA}
\newcommand*{\WM}{College of William and Mary, Williamsburg, Virginia 23187-8795}
\newcommand*{\WMindex}{37}
\affiliation{\WM}
\newcommand*{\YEREVAN}{Yerevan Physics Institute, 375036 Yerevan, Armenia}
\newcommand*{\YEREVANindex}{38}
\affiliation{\YEREVAN}
 
\newcommand*{\NOWUK}{University of Kentucky, Lexington, KY 40506}

\author{P.E.~Bosted}
\affiliation{\WM}
     \email{bosted@jlab.org}
\author {M.J.~Amaryan} 
\affiliation{\ODU}
\author {S. ~Anefalos~Pereira} 
\affiliation{\INFNFR}
\author {H.~Avakian} 
\affiliation{\JLAB}
\author {R.A.~Badui} 
\affiliation{\FIU}
\author {J.~Ball} 
\affiliation{\SACLAY}
\author {N.A.~Baltzell} 
\affiliation{\JLAB}
\affiliation{\SCAROLINA}
\author {M.~Battaglieri} 
\affiliation{\INFNGE}
\author {V.~Batourine} 
\affiliation{\JLAB}
\author {I.~Bedlinskiy} 
\affiliation{\ITEP}
% ...

\collaboration{The CLAS Collaboration}
\noaffiliation

\date{\today}


\begin{abstract}
Beam-target double-spin asymmetries and target single-spin
asymmetries were measured for the 
exclusive  $\pi^+$ electroproduction reaction
$\gamma^* p \to n \pi^+$.
The results were obtained from scattering of 6~GeV 
longitudinally polarized
electrons off longitudinally polarized protons
using the CEBAF Large Acceptance Spectrometer 
at Jefferson Lab. 
The kinematic range covered is $1.1<W<3$ GeV and $1<Q^2<6$
GeV$^2$. Results were obtained for about 6000
bins in $W$, $Q^2$, \cthcm, and $\phi^*$. Except at
forward angles, very
large target-spin asymmetries are observed 
over the entire $W$ region. Reasonable agreement is
found with phenomenological fits to previous data
for $W<1.6$ GeV, but very large differences are seen
at higher values of $W$. A GPD-based model is in poor agreement with 
the data. When combined with cross section measurements,
the present results provide powerful constraints
on nucleon resonance amplitudes at moderate and large
values of $Q^2$, for resonances with masses as high as 2.4 GeV.   
\end{abstract}
\maketitle


\section{Introduction}

\subsection{Physics Motivation}
The detailed internal
structure of the nucleon has long been studied using
exclusive electroproduction of pseudo-scalar mesons, 
a process that is 
sensitive to contributions from individual 
nucleon resonance states. Photoproduction and 
electroproduction at very
low four-momentum transfer squared ($Q^2$)  
help to determine resonance properties 
such as mass, width, parity, spin, and decay branching
ratios. Larger values of $Q^2$ are needed to study
transition form factors, and also reveal the existence
of resonances that are suppressed in photoproduction.
Initial large-$Q^2$ measurements of spin-averaged 
cross sections for exclusive $\pi^+$ electroproduction from 
Cornell~\cite{bebek76,bebek78} had limited statistical
accuracy. Recent measurements from Jefferson Lab
($\rm{JLab}$)~\cite{Horn09,HPBlok,XQian,ParkA,ParkB,ParkC} 
have greatly improved the situation. 

Experiments using polarized nucleon targets and polarized
electron beams
are particularly useful in distinguishing between
resonances of different spin, isospin, and parity, because
all single-spin asymmetries vanish in the absence of 
interference terms. This is particularly true at larger
values of $W$, where many resonances overlap. 

Nucleon resonance contributions are most important in
the central center-of-mass region ($\cos(\theta^*)=0$, or 
equivalently $t=u=s/2$). At forward angles and large $W$, 
non-resonant $t$-channel contributions dominate, and
the description of pion electroproduction is more
appropriately made using phenomenological 
Regge-pole models~\cite{regge}. 
More recently, the nuclear physics community 
has begun to evaluate exclusive electroproduction reactions 
in terms of Generalized Parton 
Distributions~\cite{JCCollins,GK09}. In such GPD models, 
spin asymmetries vanish in leading twist, and are therefore
sensitive to higher-twist operators. 

Beam asymmetries at large $Q^2$ for 
$\pi^+n$ electroproduction
from a proton target were published
from JLab for $W<1.7$ GeV \cite{ParkA}
and are also the subject of an early 
investigation for $W>2$ GeV~\cite{Avakian}. 
Beam-target asymmetries and target single-spin asymmetries
for positive and negative pions were reported from the ``eg1a'' and ``eg1b'' 
experiments at Jefferson Lab~\cite{daVita,eg1bexcl} using 1.7 to 5.7 GeV
electrons and a polarized ammonia target. The present experiment
used 6~GeV electrons only, and greatly improves the statistical 
precision of exclusive positive pion electroproduction 
asymmetries for $Q^2>1$ GeV$^2$. The present analysis closely
follows that presented in Ref.~\cite{eg1bexcl}.
After a summary of the formalism, details of the experimental
setup, analysis, and results are presented in the following
sections.

\section{Formalism}
We define the pion electroproduction cross section by:
\be
\sigma = \sigma_0 (1 + P_B A_{LU} + P_T A_{UL} + P_BP_T A_{LL}),
\ee
where $P_B$ and $P_T$ are the longitudinal beam and target polarizations,
respectively, $\sigma_0$ is the spin-averaged
cross section, and $A_{LU}$, $A_{UL}$, and $A_{LL}$ are
the beam, target, and beam-target asymmetries, respectively.
The cross sections and asymmetries are all
functions of five independent
variables. For this analysis, 
the variables $(W, Q^2, \cos(\theta^*), \phi^*, E)$
are used, where $\theta^*,\phi^*$ are the center-of-mass
decay angles of the final state with invariant mass $W$ into a meson and 
a nucleon, $Q^2$ is the squared virtual photon four-momentum,
and $E$ is the incident electron beam energy. The conventions 
used for $\theta^*$ and $\phi^*$ are given in Ref.~\cite{eg1bexcl}.
The relationship
between the present $A_{LL}$ and $A_{UL}$ observables
and the cross section components used by 
the MAID group~\cite{maid} are also given in Ref.~\cite{eg1bexcl}.

\section{Experiment}

The ``eg1-dvcs'' experiment~\cite{inclprc,dvcsprc} took data
in 2009, and had many similarities to an earlier 
experiment~\cite{eg1bexcl} which took data in 2000-2001. While
the latter experiment was designed as a broad survey in 
$W$ and $Q^2$, using beam energies from 1.6 to 5.7 GeV, the
present experiment was focused on a wide range of 
spin-dependent electroproduction reactions at large values 
of $Q^2$, using the highest available beam energy at JLab. 
Improvements
in the beam parameters, target design, detector configuration,
and data acquisition all combined to result in factors of
four to five smaller statistical uncertainties for $Q^2>1$ GeV$^2$
compared to the earlier experiment~\cite{eg1bexcl}.
A brief summary of the experimental setup is presented below: for 
more details, see Refs.~\cite{inclprc,dvcsprc}.

The present experiment used 
6 GeV longitudinally polarized electrons 
from CEBAF at JLab impinging
on a 0.025 radiation length longitudinally polarized 
solid ammonia target immersed in liquid helium 
\cite{Keith}. The target polarization
direction was along the incident electron direction,
{\it not} the direction of the momentum transfer vector.
Scattered electrons and charged pions
were detected in the CEBAF Large Acceptance 
Spectrometer (CLAS)~\cite{CLAS}.
The typical beam current was 7 nA, with a total of 
approximately 
$2\times 10^{17}$ electrons traversing the ammonia target
over the course of the experiment. 
The beam polarization, as periodically measured using
M\o ller scattering in an upstream polarimeter, averaged
85\%. 

About 90\% of the running time was on polarized
protons (NH$_3$ target), 10\% on a reference unpolarized
carbon target, and 1\% on an empty cell.
The 1.5-cm-diameter target cups contained
1 g/cm$^2$ of material immersed in a 2-cm-long
liquid helium bath.  
The sub-millimeter-diameter 
beam was slowly deflected to uniformly cover
the 1.5-cm-diameter front face of the target.
The beam position, averaged over a few minutes or longer,
was kept stable at the 0.1 mm level, using feedback from
a set of beam position monitors. 
A split superconducting solenoid magnet provided a highly
uniform 5 T magnetic field surrounding the target 
($\delta B/B \approx 10^{-5}$).

Particles were detected in CLAS for  
polar angles from 15 to 48 degrees. CLAS 
comprises six azimuthally symmetric detector
arrays embedded in a toroidal magnetic field.
Charged particle momenta and scattering angles
were measured with the drift chamber 
tracking system. 
Electrons were separated from a significantly 
larger flux of charged pions using segmented
gas Cherenkov detectors (CC, pion threshold 2.6~GeV)
and a sampling electromagnetic calorimeter (EC). 
A layer of time-of-flight scintillator counters (SC) between the
CC and EC was used for hadron identification.
The hardware
trigger system was designed to have high efficiency
for events with a scattered electron with an energy
greater than 0.3 GeV, while rejecting other events.
The hardware Cherenkov
and calorimeter thresholds were adjusted to give
a trigger rate of about 3000 Hz, with a dead time
of about 10\%. 

The standard CLAS detector set was augmented for this
experiment with an Inner Calorimeter (IC). This
calorimeter consists of an array of small lead-tungstate
crystals, each 15 cm long and roughly 2 cm square. 
The IC was not used in the present analysis, 
but blocked part of the  acceptance at small angle.

The data taking relevant to the present analysis 
was divided into two parts: Part A
(early 2009) used targets centered 
at 58 cm upstream of the CLAS center ($z_0=-58$ cm); 
Part B (mid 2009) used targets shifted an additional 10~cm upstream to 
$z_0=-68$ cm. This provided a larger acceptance for
charged particles. Combined with a higher integrated
luminosity, the bulk of the present results come from Part B.
The CLAS torus polarity was set to bend 
electrons inwards for almost all of the running time, and
the torus current was 2250 A. 
A summary of running conditions 
is given in Table~\ref{tab:parts}.
Additional information about the experimental setup
can be found in Refs.~\cite{inclprc,dvcsprc}.

\begin{table}[hbt]
\begin{tabular}{llllll}
Run Period & \> Beam Energy & \> $P_BP_T$ & \> $P_B$ \\
\hline
Part A     & \> 5.887 GeV   & \> $0.637\pm 0.011$ & \> $0.85\pm0.04$ \\
Part B     & \> 5.954 GeV   & \> $0.645\pm 0.007$ & \> $0.85\pm0.04$\\
\hline
\end{tabular}
\caption{Run period names, nominal beam energy, 
$P_BP_T$, and $P_B$, where $P_B$ ($P_T$) is
the beam (target) polarization, for the two running periods of
the experiment.
}
\label{tab:parts}
\end{table}

\section{Analysis}

\subsection{Data Processing}
A subset of the data was used to calibrate the response
of all of the CLAS detectors and instruments used
to measure beam position and current.
The alignment of the detectors, as well as the target
magnet, was also determined.

The raw data were passed through a standard CLAS analysis
package that transformed raw timing and pulse-height
signals into a set of ``particles'' for each trigger event.
Direction cosines at the 
target for charged particles, as well as their 
momenta, were determined from their tracks as 
measured by the drift chambers (DC). For neutron 
candidates, direction cosines were determined 
from their hit positions in the EC. Charged-particle 
tracks were associated
with the corresponding CC signals, EC
energy deposition, and timing from the SC using geometrical
matching. Additional details can be found in the two archival papers 
describing the eg1b inclusive analysis~\cite{inclp,incld}.

A subset of the recorded events was subsequently written to 
skimmed data files for further processing. These data files
only contained events that had a reasonable chance of
passing the event selection cuts of the present analysis.
 
\subsection{Particle Identification}
Exclusive 
$\pi^+$ electroproduction was analyzed using two topologies:
\tpI and \tpVn. Both topologies require detection of
the scattered electron and a pion. The \tpI topology
also requires the detection of a neutron.  
The total number of events passing the cuts of
topology \tpI was 32438 for Part A and 96215 for Part B.
The total number of events passing the cuts of
topology \tpV was  208835 for Part A and
 684981 for Part B.

\subsubsection{Electron identification}
Electrons were identified by requiring a signal of
at least one  photo-electron in the Cherenkov detector,
at least two thirds of the most probable electron energy 
to be deposited in the EC, and a vertex position
reconstructed within 4 cm of the nominal target center.
The electron scattering angle was required to be between
15.5 and 38 degrees. 
These cuts are not as restrictive as those placed
on electrons for the inclusive electron scattering
analysis~\cite{inclprc} of the present experiment, 
because the exclusivity
cuts discussed below remove essentially all of the
events where another type of particle might be mis-identified as
an electron. 

\subsubsection{Charged Pion Identification}
Charged pions were identified by requiring that
the time-of-arrival at the scintillator
counters be within 0.7 ns of that predicted from
the time-of-arrival of the electron in the event.
This timing cut removed all protons from the sample,
but allowed between 10\% to 100\% of $K^+$, depending
on kaon momentum. These events were removed by the
missing mass cut discussed below. Positrons were
removed from the sample by requiring small (or no)
signal in the Cherenkov detector and a 
small deposited energy in the electromagnetic calorimeter.
Also required were a vertex position
reconstructed within 4 cm of the nominal target center
and a polar scattering angle between 15 and 48 degrees.

\subsubsection{Neutron Identification}
Neutrons were identified by requiring a
deposited energy of at least 0.3 GeV in the EC, with a 
time-of-arrival at the EC corresponding to $\beta<0.95$ to
separate neutrons and photons. 
The direction cosines of the neutron were determined
from the EC hit coordinates. In some cases, the
neutrons passed through the Inner Calorimeter
on the way to the EC. Generally, this 
had no effect on the neutrons,
because the number of interactions lengths in the
IC was relatively small. In the case where the
neutron interacted in the IC, making a hadronic
shower, the exclusivity cuts on direction cosines
removed most of these events, effectively further lowering
the already low neutron detection efficiency. 
The neutron momentum
could not be determined from time-of-flight with
sufficient accuracy to be useful.

\subsection{Exclusivity Kinematic Cuts}

For both topologies, kinematic cuts
were placed to improve the signal to background ratio. The value
of kinematic cuts is two-fold. First,
most of the kinematic quantities have a wider
distribution for bound nucleons (in target materials
with $A>2$) than for free protons. Kinematic cuts therefore
reduce the dilution of the signal of interest
(scattering from polarized free protons) compared to the
background from unpolarized nucleons in materials
with $A>2$. Second, kinematic cuts are needed to
isolate single meson production from multi-meson
production and from single kaon production. 

For the \tpV topology, the only kinematic cut
available is on the missing mass. For the \tpI topology,
cuts on the cone angles of the detected neutron
further reduce nuclear backgrounds. 

\subsubsection{Electron-pion Missing Mass Cut}
For both topologies, 
the electron-pion missing mass \mxepi
should be equal to the neutron mass of 0.939 GeV.
In general, one would like the upper cut on \mxepi
to be well below  $M+m_\pi=1.08$ GeV, to avoid
contributions from multi-pion production. Placing
tighter cuts helps to reduce the nuclear background.

\begin{figure}[hbt]
\centerline{\includegraphics[width=8.7cm,angle=90]{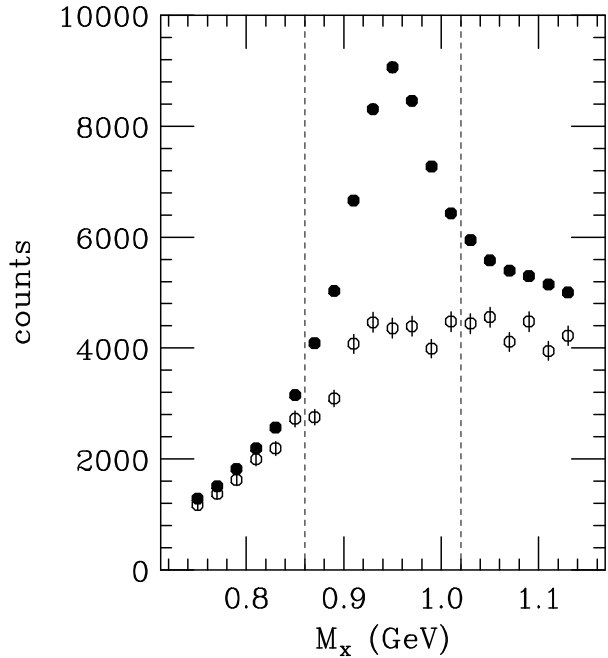}}
\caption{Sample electron-pion missing mass distribution for the
topology \tpVn, averaged over the full kinematic range of
the experiment.  Counts from the ammonia target are shown
as the solid circles and counts from the carbon target
(scaled by the ratio of integrated luminosities on 
bound nucleons) are shown as the open circles.  
The vertical dashed lines indicate the cuts used in the analysis. 
}
\label{fig:www5}
\end{figure}

The distribution in  \mxepi is shown for topology
\tpV in Fig.~\ref{fig:www5} averaged over the full
kinematic range of the experiment. 
The solid circles correspond to counts from the ammonia
target, while the open circles correspond to counts
from the carbon target, scaled by the ratio of
luminosities for $A>2$ nucleons. A clear
peak is seen near the nucleon mass from the ammonia
target, with a smaller but wider distribution from the
carbon target, that matches the wings on the ammonia
distributions on the low-mass side of the peak. On
the high side of the peak, the ammonia rates are
higher, due to the radiative tail of the single-pion
production, and the gradual turn-on of multi-pion
production. The vertical dashed lines show the
cuts used: $0.86<M_x^{e\pi}<1.02$ GeV. Within the
cut region, approximately half of the events come
from nucleons in nuclei with $A>2$, and half from free protons. 

\begin{figure}[hbt]
\centerline{\includegraphics[width=8.7cm,angle=90]{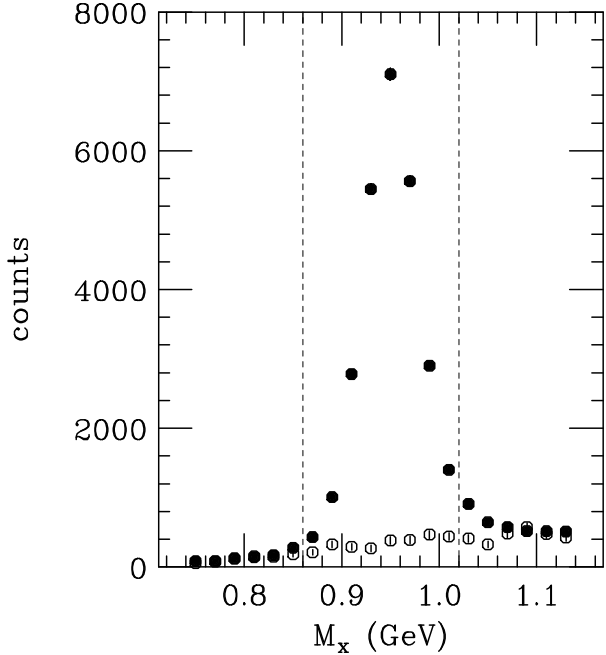}}
\caption{Same as Fig.~\ref{fig:www5}, but for the topology
\tpIn. Cuts on the neutron angle have been applied.
}
\label{fig:www1}
\end{figure}

The distribution in  \mxepi is shown for topology
\tpI in Fig.~\ref{fig:www1}. The nuclear background is
greatly reduced in this case, because additional
cuts can be placed on the direction cosines of the
detected neutron.

The spectra were examined to
see if the optimal cut value depends on $W$,
$Q^2$, \cthcm, or \phicm. Although the peak widths
vary somewhat with kinematic variables, a constant cut value did not
degrade the signal to noise ratios by more than a few percent. 

\subsubsection{Neutron Angular Cuts}
\begin{figure}[hbt]
\centerline{\includegraphics[width=8.7cm,angle=90]{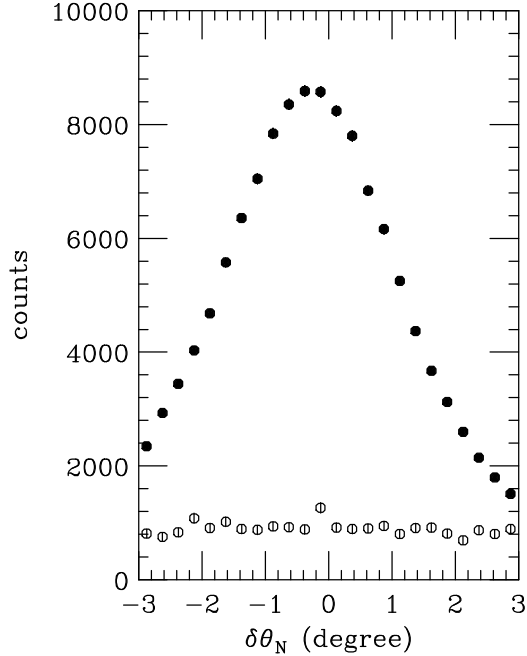}}
\caption{Distribution of the in-plane angular difference in
predicted and observed nucleon direction cosines 
for the topology \tpIn. 
The black points are for the ammonia target, while
the open circles are from the carbon target, scaled
by integrated luminosity. The analysis cuts correspond
to the edge of the histogram.
All other relevant exclusivity cuts (i.e. on \mxepi and 
$\delta \phi_N$) have been applied.
}
\label{fig:dthn}
\end{figure}

\begin{figure}[hbt]
\centerline{\includegraphics[width=8.7cm,angle=90]{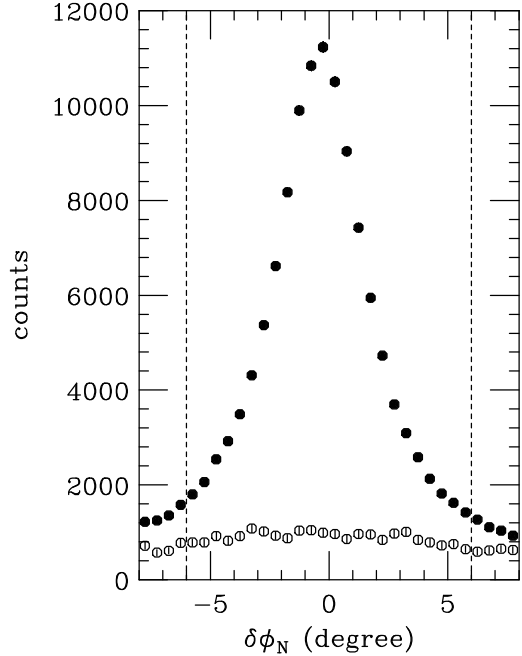}}
\caption{Same as Fig.~\ref{fig:dthn}, except now for
the out-of-plane angular difference (after application of
cuts on \mxepi and $\delta \theta_N$) 
The vertical dashed lines
indicate the cuts used in the analysis. 
}
\label{fig:dphin}
\end{figure}

For the topology \tpIn, cuts on the cone angles of the neutron
are very useful in 
rejecting background from $A>2$ materials in the target. From the
kinematics of the detected electron and pion, the
direction cosines of the recoil neutron are calculated,
and compared with the observed angles. We denote
the difference in predicted and observed angles
as $\delta \theta_N$ in the in-plane direction and
$\delta \phi_N$ in the out-of-plane direction (which
tends to have worse experimental resolution). Distributions
of these two quantities are shown
in Figs.~\ref{fig:dthn} and \ref{fig:dphin},
respectively. It can be seen that with cuts on $M_x$ and
the complementary angle, the nuclear background is relatively small 
and flat compared to the peaks from the free proton.  
We used the cuts $\lvert \delta \theta_N \rvert <3^\circ$
and $\lvert \delta \phi_N \rvert <6^\circ$ for all kinematic bins. Events
that failed either one of these cuts were
not moved over to the \tpV topology event sample. 

\subsection{Kinematic Binning}\label{sec:kin}

\begin{figure}[hbt]
\centerline{\includegraphics[width=8.7cm,angle=90]{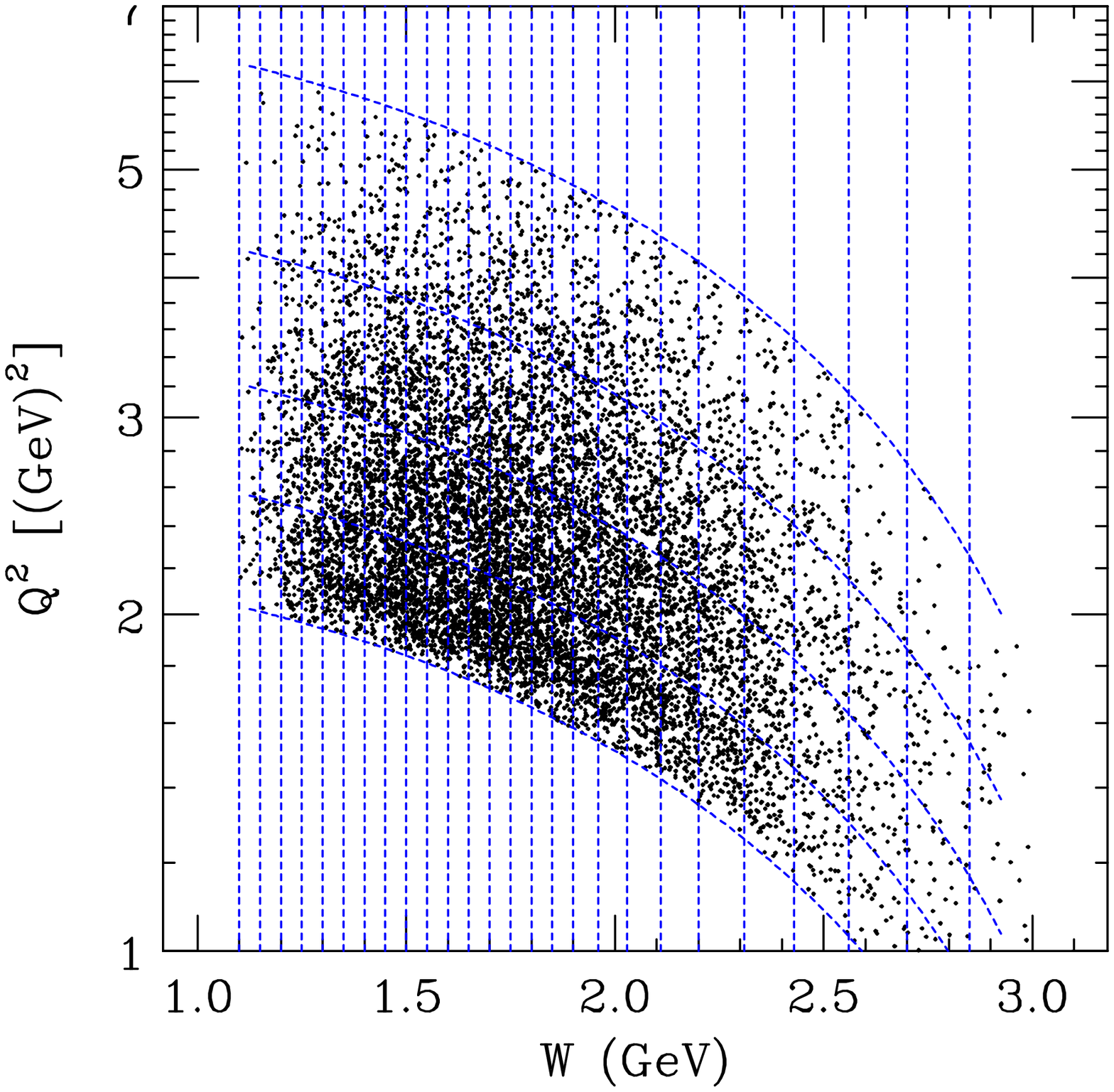}}
\caption{Distribution in $(W,Q^2)$ of events for the \tpV
topology passing all exclusivity cuts. The vertical
dashed lines show the limits of the $W$ bins used in
the analysis, while the left-to-right curves show
the bin limits in $Q^2$, defined by fixed bins in 
$\theta_e$ of 15.5, 18, 21, 26, and 38 degrees (from bottom to top).
}
\label{fig:wq2}
\end{figure}

The kinematic
range of the experiment is $1.1<W<3$ GeV and 
$1<Q^2<6$ GeV$^2$. As shown in Fig.~\ref{fig:wq2},
the range in $Q^2$ changes with $W$. We therefore
made four bins in $Q^2$, where the
limits correspond to electron scattering angles
of 15.5, 18, 21, 26, and 38 degrees. In order to
study possible resonance structure, we 
used fixed $W$ bins of width 0.05 GeV for $W<1.9$ GeV,
which is comparable to the experimental 
resolution. For $W>1.9$ GeV, the bin widths gradually
increase, to achieve roughly equal counting rates, with
bin boundaries at 1.90, 1.96, 2.03, 2.11, 2.20, 2.31, 2.43, 2.56,
2.70, 2.85 and 3 GeV.        
The bin limits are shown in Fig.~\ref{fig:wq2}.

An examination of event rates showed a strong forward
peaking in \cthcmsp for both topologies
studied, roughly independent of $(W,Q^2)$. There are
essentially no events with $\cos(\theta^*)<-0.2$.
We decided to use six bins in $\cos(\theta^*)$, with
boundaries at -0.2, 0.2, 0.44, 0.63, 0.78, 0.9, and 0.995.
The upper-most boundary of 0.995 was chosen instead of 1.0
because the average
resolution in \phicmsp becomes larger than 30 degrees above
$\cthcmn=0.995$, making it increasingly problematic to determine the
$\phi^*$-dependence of the spin asymmetries at very forward angles.
We used 12 bins in $\phi^*$, equally spaced
between 0 and $2\pi$. 

A strong consideration in choosing the bin sizes was that
we required at least ten counts in a given bin in order
to have approximately Gaussian statistical uncertainties. 
The total number of bins is 7488, of which about 6000
had enough events to be included in the final results.

\newpage\section{Asymmetries}
Spin asymmetries were formed as follows:
\be
A_{LL} = \frac{
   N^{\uparrow\downarrow} + 
   N^{\downarrow\uparrow} -
   N^{\uparrow\uparrow} - 
   N^{\downarrow\downarrow}}{
   N_{tot} \hskip .05in f \hskip .05in P_BP_T},
\ee
\be
A_{UL} = \frac{
   N^{\uparrow\uparrow} + 
   N^{\downarrow\uparrow} -
   N^{\uparrow\downarrow} - 
   N^{\downarrow\downarrow}}{
   N_{tot} \hskip .05in f \hskip .05in P_T},
\ee
where the symbols $N$ represent the number of events
in a given helicity configuration, divided by the
corresponding integrated beam current. The first superscript
refers to the beam polarization direction and the
second to the target polarization direction. The total
number of counts is denoted by
$N_{tot}=N^{\uparrow\uparrow} + 
               N^{\downarrow\uparrow} +
               N^{\uparrow\downarrow} + 
               N^{\downarrow\downarrow}$
and $f$ is the dilution factor, defined as the fraction
of events originating from polarized free protons, compared
to the total number of events.

\subsection{Beam and Target Polarization}
The product of beam 
polarization ($P_B$) and
target polarization ($P_T$) was
determined using the well-understood beam-target spin
asymmetry in elastic $ep$ scattering. The results
are listed in Table~\ref{tab:parts}. The beam polarization
was measured using M\o ller scattering, and is also
listed in the table. The proton target polarization was 
determined by dividing $P_BP_T$ by $P_B$. This proved
to be more accurate than using direct NMR measurements
of the target polarization, which were relatively
accurate from run-to-run, but had a large overall
normalization uncertainty. 

\subsection{Dilution Factor}
\label{sec:df}
The dilution factor $f$ is defined as the ratio of 
scattering rate from free nucleons to the scattering 
rate from all nucleons in the target.
With the assumption that the cross section per nucleon
is the same for bound protons in all of the nuclear
materials (with $A>2$) in a given target, and also that
the effective detection efficiency is the same for the
ammonia and carbon targets, then
\be
f = 1 - R_{A>2} \frac{N_C}{N_{NH_3}},
\label{Eq:f}
\ee
where 
$N_C$ and $N_{NH_3}$ are the number of counts from the
carbon and ammonia targets respectively,
measured in a given kinematic bin for a given topology,
normalized by the corresponding integrated beam charge. The
symbol $R_{A>2}$ denotes the ratio of the number of bound nucleons
in the  ammonia target to the number of bound nucleons in the
carbon target. Bound nucleons are defined to be in
materials with atomic number $A>2$.
The latter was determined from a detailed analysis of the
target composition using inclusive electron scattering
rates from ammonia, carbon, and 
empty targets, yielding
$R_{A>2}=0.71$ for Part A and $R_{A>2}=0.72$ for Part B.

Because the integrated luminosity on the carbon target
was about ten times lower than on the ammonia
target, there is a large amplification of the uncertainty
on the ratio of carbon to ammonia counts, 
$\frac{N_C}{N_{NH_3}}$. In many cases, this would lead
to unphysical values of $f$ (i.e. $f<0$). We therefore
took advantage of the fact that $f$ is a very slowly
varying function of kinematic variables, and did a global fit
to $\frac{N_C}{N_{NH_3}}$. 
The fit values were then
used to evaluate $f$ in each kinematic bin.

As in Ref.~\cite{eg1bexcl}, the functional forms for the fit 
contained 25 terms of the
form $p_i \cos^{N_c}(\theta^*) W^{N_W} (Q^2)^{N_Q}$, where 
$p_i$ is a free parameter, and the exponents $N_C$, $N_W$, and
$N_Q$ range from 0 to 3 (although not all possible terms were
included). An additional eight terms were included to 
account for the influence
of the three prominent nucleon resonances centered at 
$1.23$ GeV, 
$1.53$ GeV, and 
$1.69$ GeV,
with widths 
$0.220$ GeV, 
$0.120$ GeV, and
$0.120$ GeV.
The reason that these resonance terms are
needed is that the nucleon resonances are effectively broadened
in the target materials with $A>2$ by Fermi motion.
This generates resonant-like structures in
the ratio of carbon to ammonia count rates. 
Tests were made to see if any $\phi^*$-dependent
terms would improve the fits. No significant 
improvements were found. 

\begin{figure}[hbt]
\centerline{\includegraphics[width=8.7cm,angle=90]{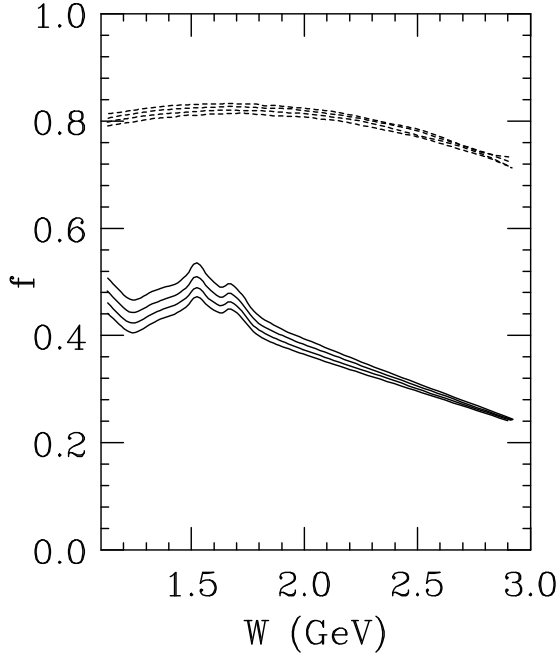}}
\caption{Dilution factors as a function of $W$ for
the \tpI topology (dashed curves) and the \tpV topology
(solid curves) for the four $Q^2$ bins of this experiment and
a typical bin in $\cos(\theta^*)$.
}
\label{fig:dilff1}
\end{figure}

The dilution factors for Part B for 
the two topologies are shown in 
Fig.~\ref{fig:dilff1} as a function of $W$ for the four $Q^2$
bins of this analysis and a typical bin in \cthcm. 
 For the fully
exclusive topology, \tpIn, the dilution factor
is large, about 0.8 on average, corresponding
to the good rejection of background that is possible
with the exclusivity cuts when the recoil neutron
is detected.  For the topology \tpVn, the dilution 
factor is reasonably good for $W<2$ GeV,
averaging about 0.45, with significant resonant 
structure visible. For $W>2$ GeV, there is a
trend for $f$ to decrease, dropping to values as 
low as 0.25 at the highest values of $W$. This is
because Fermi broadening results in an increasing amount
 of multi-pion production from
the nuclear target materials. The $Q^2$-dependence is 
relatively weak for both topologies.
Because Part A had much lower statistical accuracy than
Part B, we used the Part B fits for Part A.

\subsection{Combining Data Sets}
The entire asymmetry analysis was performed separately
for Part A and Part B.
The results were combined
by averaging asymmetries, weighted by their 
respective statistical uncertainties, for each of the
4-dimensional bins. Since the two configurations
differ only in the acceptance function, which should
cancel in forming the asymmetries, the expectation is
that they should be fully compatible statistically.
This expectation was verified for 
both asymmetries for all three topologies. 

\subsection{Combining Topologies}

\begin{figure}[hbt]
\centerline{\includegraphics[width=13cm,angle=90]{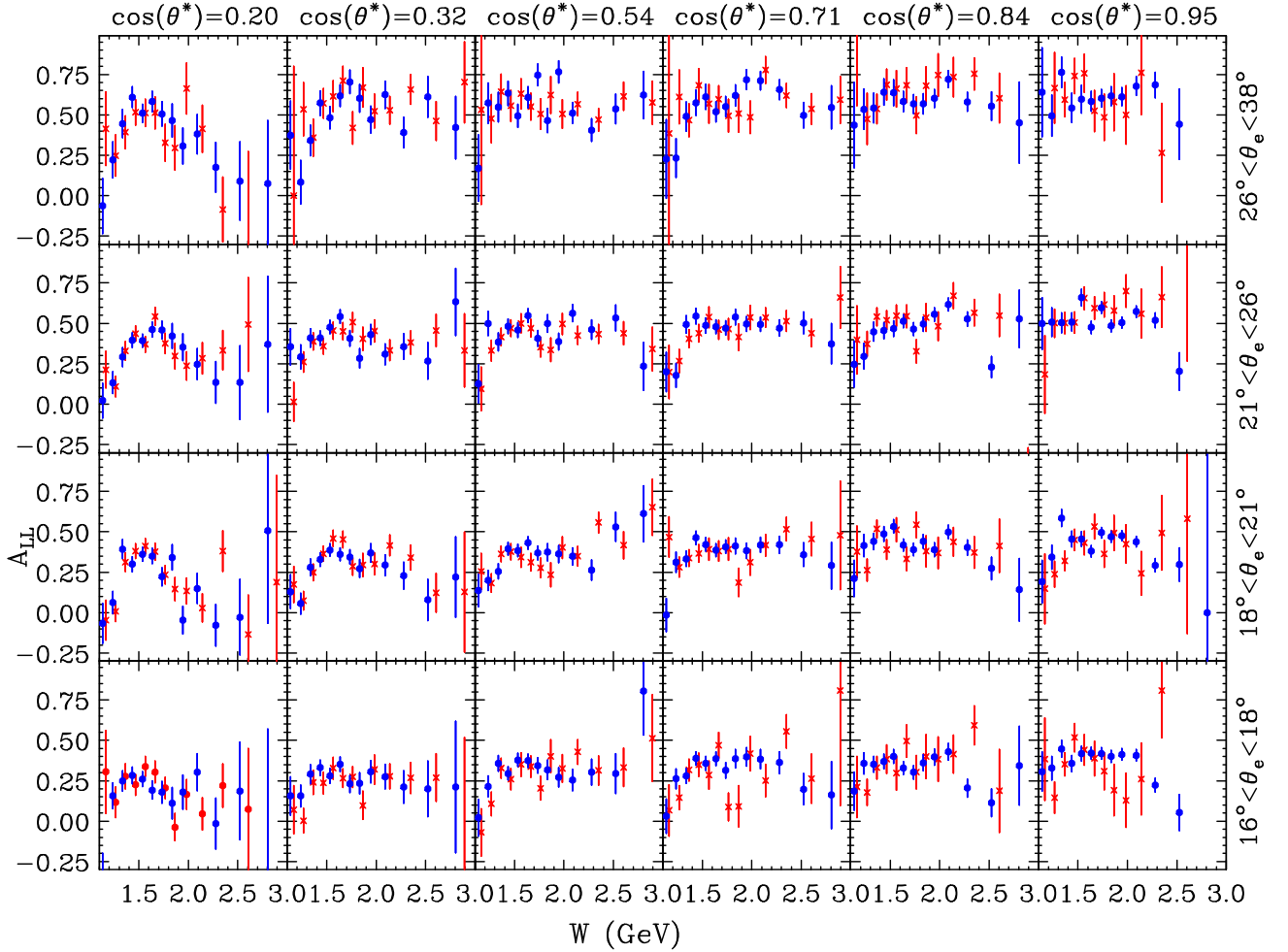}}
\caption{(color online) Beam-target double spin asymmetry $A_{LL}$
for the \tpI topology (red crosses) compared to the 
\tpV topology (blue circles),
averaged over $\phi^*$,
 as a function of $W$, in
the six \cthcmsp bins of this analysis
and the four $Q^2$ ($\theta_e$) 
bins used. 
}
\label{fig:ALLpip2}
\end{figure}

The next step was to combine the fully exclusive topology with the one
with a missing neutron. 
For both  asymmetries, the
topologies were found to be statistically compatible. 
This good agreement between topologies can be observed
 by visual examination of plots in which both topologies
are plotted together, such as 
Fig.~\ref{fig:ALLpip2},
which show $A_{LL}$ for the two $\pi^+$ topologies 
as a function of $W$ in a grid over
$\theta_e$ (i.e. $Q^2$) and $\cos(\theta^*)$. 
In this
figure, adjacent bins in $W$ were averaged together
and a straight average over $\phi^*$ was performed. 

\subsection{Radiative Corrections}
Radiative corrections take into
account that the incident beam energy, scattered
electron energy, or the electron scattering angle
at the vertex can all be different from those
measured in the detector, due to electrons 
radiating photons  in the field of 
a nucleon or nucleus. Although the corrections are 
significant for spin-averaged exclusive cross sections, 
they are negligible for spin asymmetries, due to the
facts that Bremsstrahlung is largely spin-independent,
and the cross section variation is small within the
exclusivity cuts used for a given kinematic bin. This
was verified by explicit calculations using the
Mo-Tsai formalism~\cite{tsai} with the equivalent
radiator approximation (internal radiation equivalent
to external radiation) and the angle peaking approximation
(photon emitted along the incident or scattered electron
direction only). In these calculations, we used
the MAID fit~\cite{maid} to describe the 
cross section and asymmetry variations within each
kinematic bin. The calculations were performed using
a Monte Carlo simulation technique. Within the
statistical uncertainty of the calculation (typically
$\delta A=0.005$ for a given kinematic bin), no
significant deviations from zero were observed. 
The average depolarization of the electron  from
Bremsstrahlung was also evaluated and found to be
much less than 1\%. 

\subsection{Polarized Nitrogen Correction}
As is discussed in Ref.~\cite{inclprc}, the nitrogen
in the ammonia targets is slightly polarized, and
in the case of inclusive electron scattering, a 
correction of about 1.8\% to the beam-target
asymmetry is needed. In the present exclusive analysis,
the correction is reduced to about 0.5\% for \tpV and
less than 0.2\% for \tpIn, 
because most of the events from nitrogen are removed
by the exclusivity cuts. No corrections were applied in the
present analysis, and this omission is accounted for in the
systematic uncertainty budget.

\subsection{Systematic Uncertainties}
The dominant systematic uncertainty on all the asymmetry
results is an overall scale uncertainty from the
beam and target polarizations. 
The uncertainty in $A_{LL}$ is relatively small (1.4\%)
because $P_BP_T$ was well-measured
using $ep$ elastic scattering. The relative 
uncertainty in $A_{UL}$ is larger (4\%) due to the
uncertainty in $P_B$, from which we obtained $P_T$
by dividing $P_BP_T$ by $P_B$.

The other source of normalization uncertainty
is the dilution factor. As discussed in more detail
in Ref.~\cite{inclprc}, the uncertainties in the target
composition correspond to about a 2.5\% relative uncertainty in the 
amount of background subtraction, which corresponds to
1\% to  1.5\% in the asymmetry results, for the 
missing neutron topology, and less than 0.5\%
for the fully exclusive topology.

Another source of systematic uncertainty is in the 
factor  $R_{A>2}$. We compared
three methods of determining this factor: a study of
inclusive electron scattering rates; fits to the low
electron-pion missing mass spectra; and the value that gives
the best agreement for $A_{LL}$ between the fully
exclusive topology and the topology where the
recoil nucleon is not detected. This last technique
relies on the fact that the fully exclusive topology
has much less nuclear background. From these comparisons,
we estimate a systematic uncertainty of about 2\% (relative)
for $R_{A>2}$. This translates
into approximately 1.5\% (at low $W$) to 2.5\%  (at high $W$)
overall normalization uncertainties on
both $A_{LL}$ and $A_{UL}$.
 
It is also possible for assumptions made in the 
dilution factor fitting, such as the lack of $\phi^*$
dependence, to result in point-to-point systematic
uncertainties. Based on trying out several different functional
forms to the fit, these were found to be much smaller
than the point-to-point statistical uncertainties.

Finally, it is clear from Fig.~\ref{fig:www5} that
the cut on electron-pion missing mass is not 100\%
effective at removing multi-pion production for
the topology with one missing nucleon. 
Since the contamination is larger for
$M_x^{e\pi}>M$ than for $M_x^{e\pi}<M$, we 
divided the data
into two distinct sets, based on the above criteria,
and compared both $A_{LL}$ and $A_{UL}$ asymmetries. 
We obtained $\chi^2$/d.f.=0.98 ($\chi^2$/d.f.=1.02) 
for agreement of the two $A_{LL}$ ($A_{UL}$) data sets,
indicating that the admixture of some multi-pion
events into the single pion samples does not affect
the final asymmetry results significantly.

Adding the above sources of uncertainty in quadrature, we 
obtain an overall normalization uncertainty of 3\% for 
$A_{LL}$ and 5\% for $A_{UL}$.

\section{Results}

With over 7000 kinematic points, each with relatively large
uncertainties, it is a challenge to portray the entire data set in
a meaningful way. For plotting purposes,
we therefore averaged together adjacent bin
triplets or quartets in $W$ and adjacent bin pairs in $Q^2$. 
The complete set of results is available in the
CLAS physics data base~\cite{clasdb} and in the 
Supplemental Material associated with this article~\cite{SMp}.
All results are for the fully 
exclusive topology and the topology with a missing neutron
combined together, as explained above.

\subsection{$A_{LL}$}
\begin{figure}[hbt]
\centerline{\includegraphics[width=13cm,angle=90]{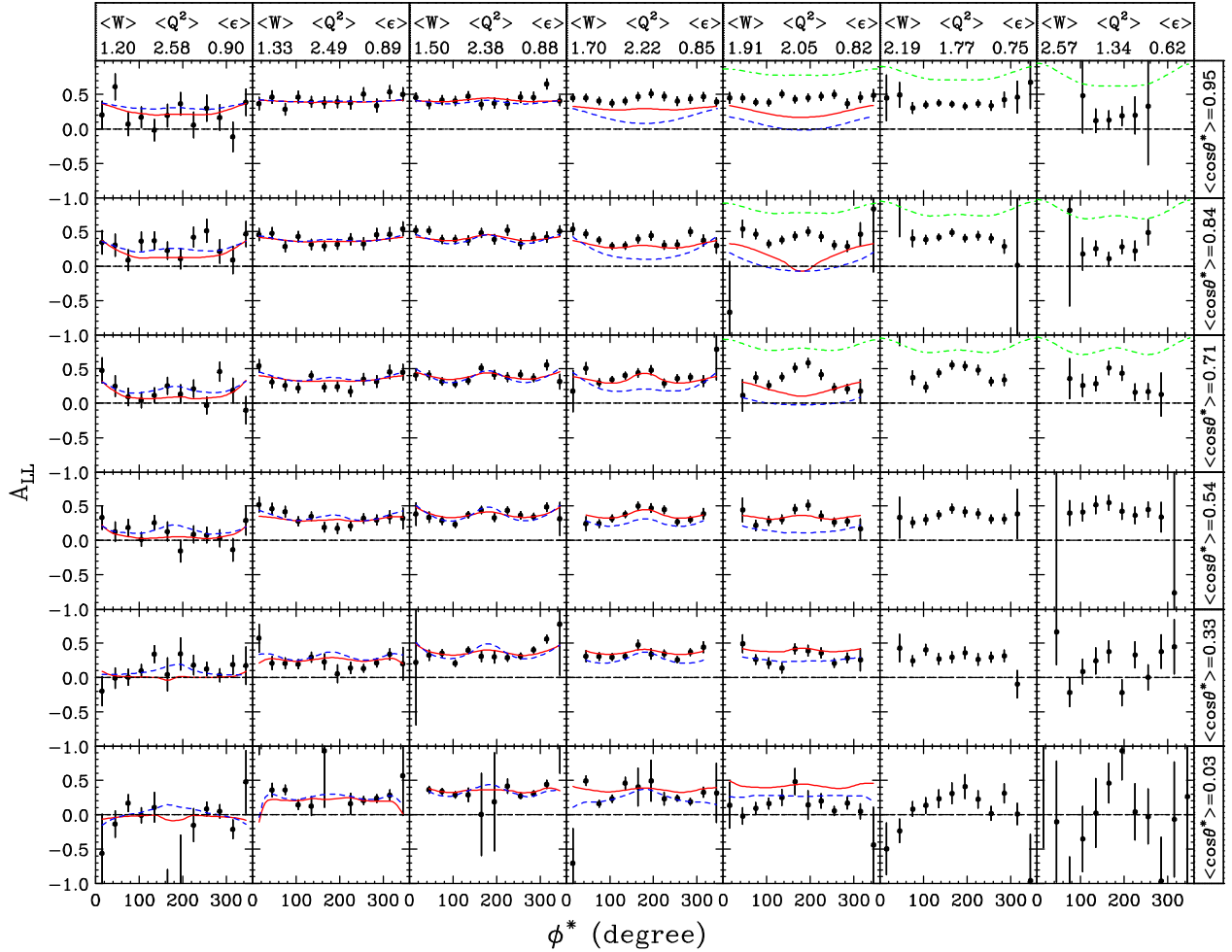}}
\caption{(color online) Beam-target double spin asymmetry $A_{LL}$
for the reaction \tpI as a function of  $\phi^*$
in seven bins in $W$ (columns) and 
six \cthcmsp bins (rows). The
results are from the two lower $Q^2$ bins of this analysis.
The error bars reflect statistical uncertainties only.
The solid red curves are from the MAID 2007 fit~\cite{maid}, 
the blue long-dashed curves are from a JANR fit~\cite{janr},
and the green short-dashed curves are for the 
GPD-inspired model from
Goloskokov and Kroll~\cite{GK09}.
}
\label{fig:ALLloq}
\end{figure}

\begin{figure}[hbt]
\centerline{\includegraphics[width=13cm,angle=90]{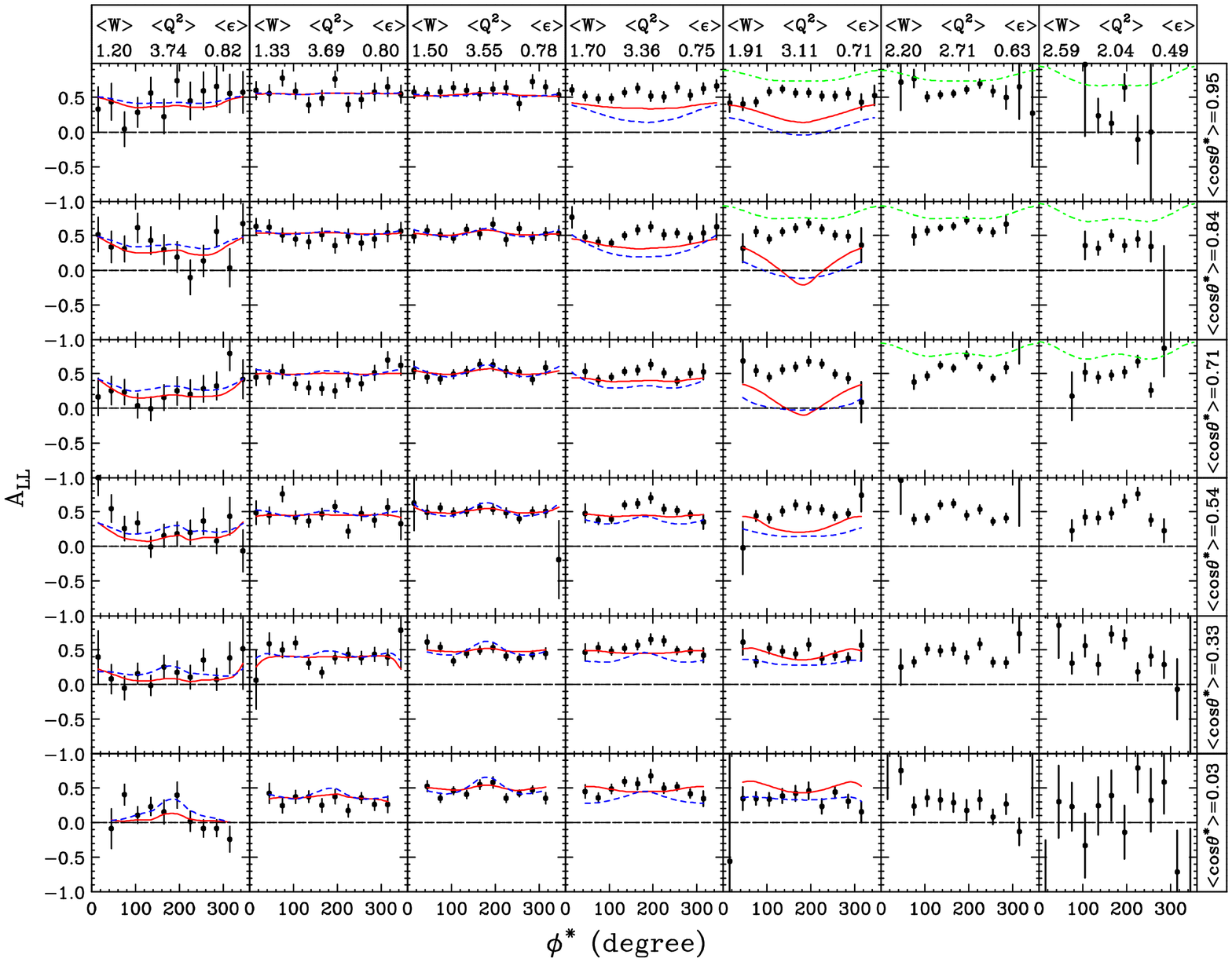}}
\caption{Same as Fig.~\ref{fig:ALLloq}, except for the
 two larger $Q^2$ bins of this analysis.
}
\label{fig:ALLhiq}
\end{figure}

The results for the beam-target spin asymmetry 
$A_{LL}$ are plotted as a function of
$\phi^*$ in seven bins in $W$ and six bins in
$\cos(\theta^*)$ in Fig.~\ref{fig:ALLloq} for the lower
$Q^2$ data and in Fig.~\ref{fig:ALLhiq} for the higher
$Q^2$ data. There is very little difference between
these plots, indicating a weak dependence
on $Q^2$ for a given kinematic bin.

The main feature of the data is a
relatively large and positive asymmetry (averaging about
0.4) for most kinematic bins. 
The major exception is
for the lowest $W$ bin, centered on the $\Delta(1232)$
resonance, where the values of $A_{LL}$ are closer to zero.
This feature is expected because the $\Delta(1232)$
transition is dominated by spin-1/2 to spin-3/2, which
gives a negative value of $A_{LL}$, balancing the
positive contribution from the Born terms. Of particular
interest are the bins centered on $W=1.70$, $W=1.91$  and
$W=2.19$ GeV. Here, $A_{LL}$ is roughly 0.4, independent
of $\phi^*$, at forward angles where $t$-channel processes
dominate. At lower values of $\cos(\theta^*)$, an
increasingly large $\phi^*$-dependence can be seen,
with a noticeable enhancement near $\phi^*=180^\circ$.
This suggests the importance of $s$-channel
resonance excitations. 

Also shown on the plots are the
results of two representative fits to previous data (limited
to $W<2$ GeV):
the 2007 version of the
MAID unitary isobar fit~\cite{maid} and
the Unitary Isobar version of the JLab Analysis
of Nucleon Resonances (JANR) fit~\cite{janr}, averaged with
the same weighting as the data points. Formally, these two
fits are rather similar in nature, but differ in the data
sets used and in the functional forms used for the 
$Q^2$-dependence of the resonance form factors. 
By and large, both the MAID 2007 and the JANR fits describe
the data reasonably well up to $W=1.6$ GeV, 
with large differences in
the $\phi^*$-dependence appearing at larger $W$. 
Also shown on the plots are the GPD-based model of
Goloskokov and Kroll~\cite{GK09}, which has no explicit
$s$-channel resonance structure included. This model 
generally predicts larger values of $A_{LL}$ than observed.

\subsection{$A_{UL}$}
\begin{figure}[hbt]
\centerline{\includegraphics[width=13cm,angle=90]{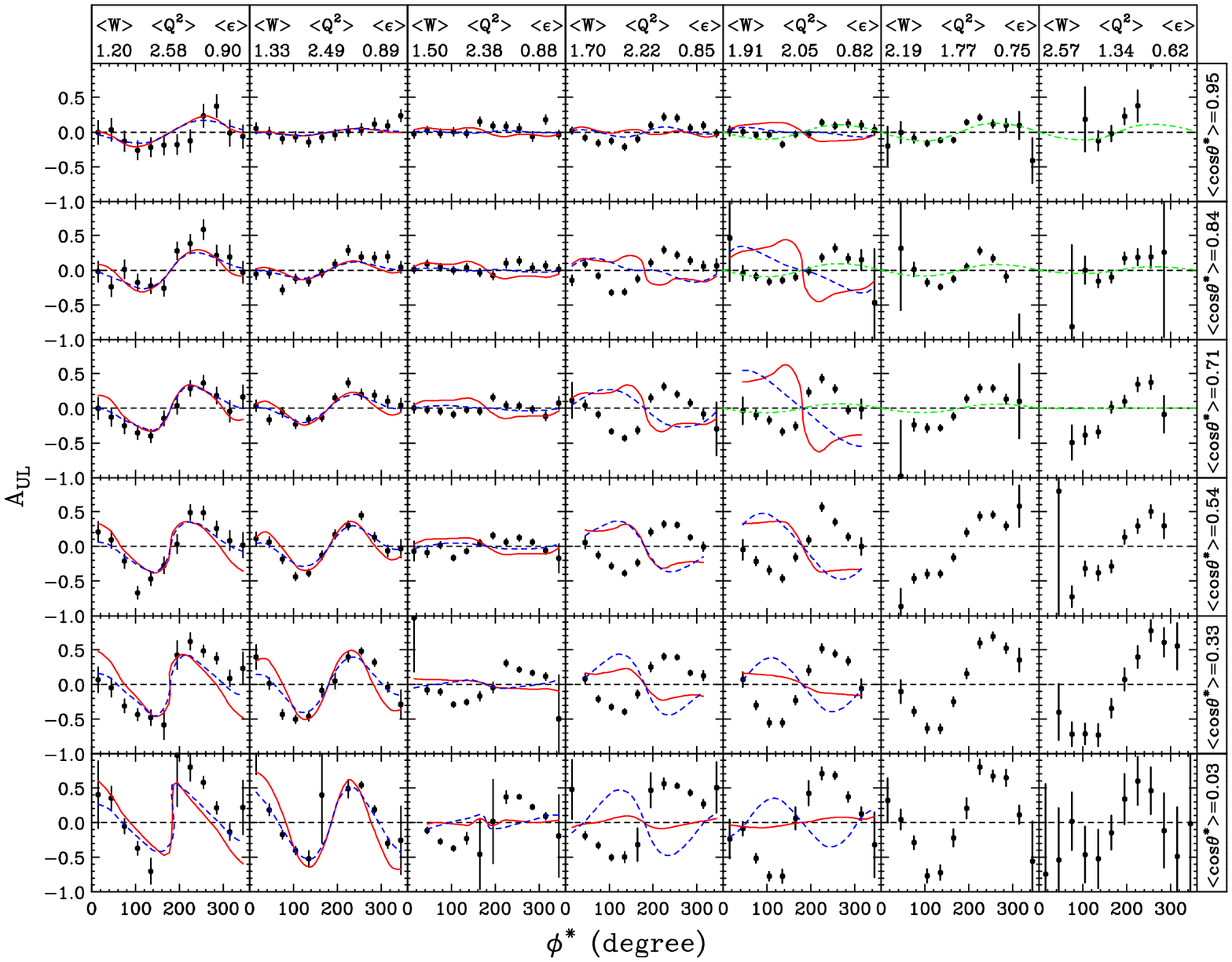}}
\caption{Target single-spin asymmetry $A_{UL}$
for the reaction \tpI as a function of  $\phi^*$
in seven bins in $W$ (columns) and 
six \cthcmsp bins (rows). The
results are from the two lower $Q^2$ bins of this analysis.
The error bars reflect statistical uncertainties only.
The solid red curves are from the MAID 2007 fit~\cite{maid}, 
the blue long-dashed curves are from a JANR fit~\cite{janr},
and the green short-dashed curves are for the 
GPD-inspired model from
Goloskokov and Kroll~\cite{GK09}.
}
\label{fig:AULloq}
\end{figure}

\begin{figure}[hbt]
\centerline{\includegraphics[width=13cm,angle=90]{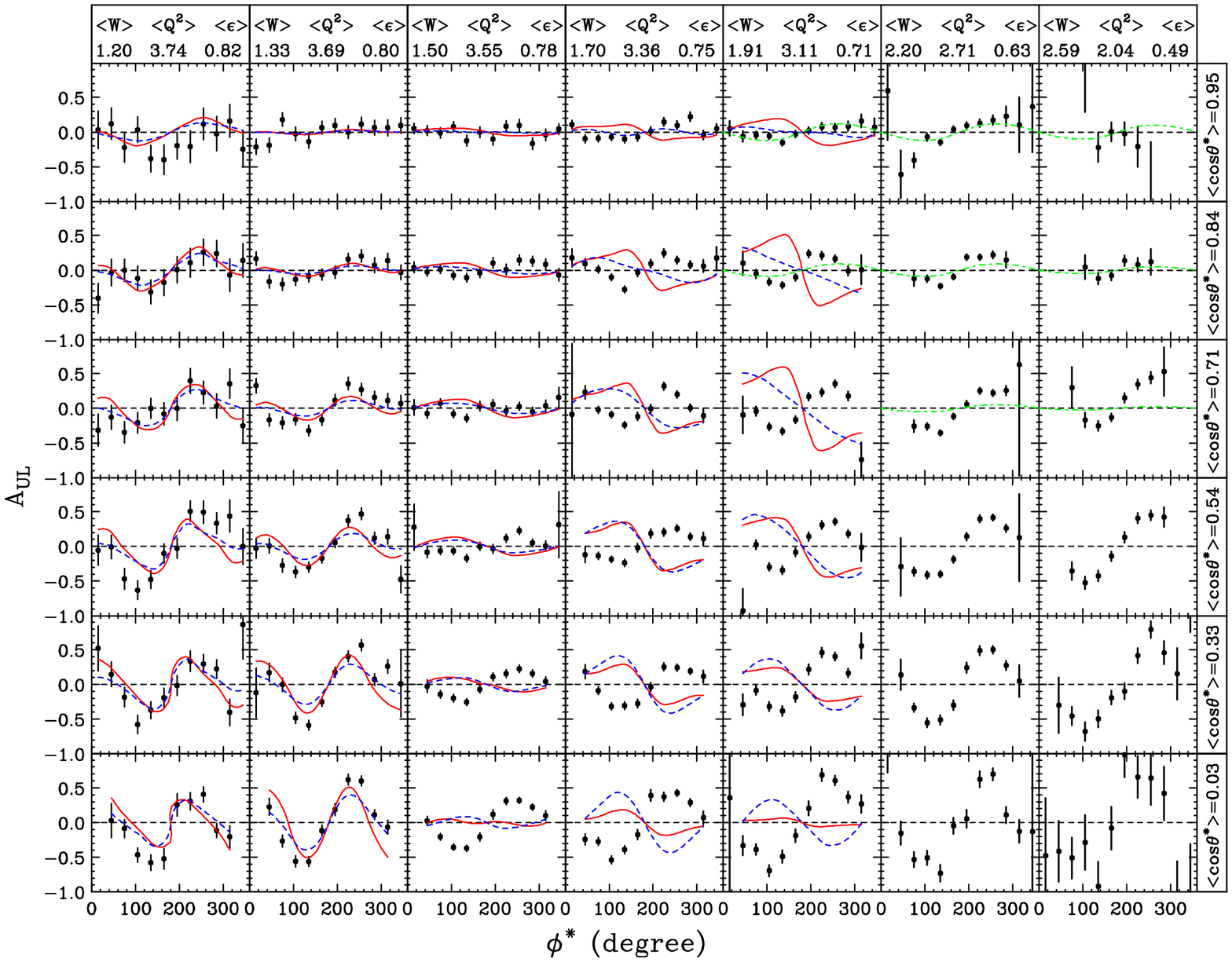}}
\caption{Same as Fig.~\ref{fig:AULloq}, except for the
 two larger $Q^2$ bins of this analysis.
}
\label{fig:AULhiq}
\end{figure}

The results for the target spin asymmetry 
$A_{UL}$ are plotted as a function of
$\phi^*$ in seven bins in $W$ and six bins in
$\cos(\theta^*)$ in Fig.~\ref{fig:AULloq} for the lower
$Q^2$ data and in Fig.~\ref{fig:AULhiq} for the higher
$Q^2$ data. It can be seen that the $Q^2$-dependence
of the results is weak. 
The main feature of the data is a positive
$\sin(\phi^*)$ modulation that is small at forward
angles, and grows to nearly maximal values at
central angles, even at the largest values of $W$.

The sign and magnitude of this
modulation is well reproduced by the MAID and JANR fits
for $W<1.4$ GeV,
where the $\Delta(1232)$ resonance dominates. At 
larger values of $W$, both fits predict a sign
change in the $\sin(\phi^*)$ modulation, which is not
observed in the data. The magnitude of the modulation
is also much larger in the data than in the previous fits
near $\cos(\theta^*)=0$. The GPD-inspired model from
Goloskokov and Kroll~\cite{GK09} agrees well with the small
asymmetries observed at very forward angles, but does not predict
the large asymmetries observed at smaller values of 
$\cos(\theta^*)$. 

Combined with the results
for $A_{LL}$, the results for $A_{UL}$ strongly suggest
that there are important nucleon resonance contributions
to exclusive pion electroproduction for $W>1.6$ GeV
and $Q^2>1$ GeV$^2$. For example, the Particle
Data Group~\cite{pdg} lists four ``3-star'' and ``4-star'' $N^*$ resonances
with masses above 2 GeV (at 2190, 2220, 2250, and 2600 MeV)
and a ``4-star'' $\Delta$ resonance with mass 2420 MeV.

\section{Summary}
Beam-target double-spin asymmetries and target single-spin
asymmetries were measured for the 
exclusive  $\pi^+$ electroproduction reaction
$\gamma^* p \to n \pi^+$.
The results were obtained from scattering of 6~GeV 
longitudinally polarized
electrons off longitudinally polarized protons
using the CEBAF Large Acceptance Spectrometer 
at Jefferson Lab. 
The kinematic range covered is $1.1<W<3$ GeV and $1<Q^2<6$
GeV$^2$. Results were obtained for about 6000
bins in $W$, $Q^2$, \cthcm, and $\phi^*$. Except at
forward angles, very
large target-spin asymmetries are observed 
over the entire $W$ region. 
Reasonable agreement is
found with the phenomenological MAID 2007 fit~\cite{maid}
and the 2009 JANR fit~\cite{janr} to previous data
for $W<1.5$ GeV, but very large differences are seen
at higher values of $W$, where no large-$Q^2$ 
data were available when the fits
were made. The large target-spin asymmetries
are also not accounted for by a GPD model.  We anticipate
that the present target and beam-target asymmetry data, when
combined with beam-spin asymmetry and spin-averaged cross
section data in new global analyses, will yield major insights 
into the structure of the proton and its many excited states.

\section*{Acknowledgments}
We thank I. Aznauryan for providing the JANR source
code and L. Tiator for providing the MAID 2007
source code. 
We thank X. Zheng for suggesting the functional form of
the dilution factor fit.
We acknowledge the outstanding efforts of the staff
of the Accelerator and the Physics Divisions at Jefferson Lab that made
this experiment possible.  This material is based
upon work supported by the U.S. Department of Energy, 
Office of Science, Office of Nuclear Physics under contract 
DE-AC05-06OR23177 and the National Science Foundation. 
Partial support was provided by 
the Scottish Universities Physics Alliance (SUPA),
the United Kingdom's Science and Technology Facilities Council,
the National Research Foundation of Korea,
the Italian Instituto Nazionale di Fisica Nucleare, the French Centre
National de la Recherche Scientifique, and the French Commissariat \`{a}
l'Energie Atomique.
The Southeastern Universities Research Association (SURA) operates
the Thomas Jefferson National Accelerator Facility for the
United States Department of Energy under contract DE-AC05-84ER-40150.

\end{document}